\documentclass{aastex631}
\usepackage{braket}
\usepackage{amsmath}

\received{September 1, 2023}
\accepted{February 5, 2024}
\submitjournal{ApJ}

\begin{document}

\title{Comparing Observed with Simulated Solar Disk Center Scattering Polarization in the Sr~{\sc i} 4607~\AA \ line}

\correspondingauthor{Franziska Zeuner}
\email{zeuner@irsol.ch}

\author[0000-0002-3594-2247]{Franziska Zeuner}
\affiliation{Istituto ricerche solari Aldo e Cele Daccò (IRSOL), Faculty of Informatics, Università della Svizzera italiana, CH-6605 Locarno, Switzerland}
\affiliation{ Max-Planck-Institut f{\"u}r Sonnensystemforschung, Justus-von-Liebig-Weg 3, D-37077 G{\"o}t\-tingen, Germany}

\author[0000-0003-1465-5692]{Tanaus\'{u} del Pino Alem\'{a}n}
\affiliation{Instituto de Astrof\'{i}sica de Canarias, E-38205 La Laguna, Tenerife, Spain}
\affiliation{Departmento de Astrof\'{i}sica, Facultad de F\'{i}sica, Universidad de La Laguna, E-38206 La Laguna, Tenerife, Spain}

\author[0000-0001-5131-4139]{Javier Trujillo Bueno}
\affiliation{Instituto de Astrof\'{i}sica de Canarias, E-38205 La Laguna, Tenerife, Spain}
\affiliation{Departmento de Astrof\'{i}sica, Facultad de F\'{i}sica, Universidad de La Laguna, E-38206 La Laguna, Tenerife, Spain}
\affiliation{Consejo Superior de Investigaciones Cientif\'{i}cas, Spain}

\author[0000-0002-3418-8449]{Sami K. Solanki}
\affiliation{ Max-Planck-Institut f{\"u}r Sonnensystemforschung, Justus-von-Liebig-Weg 3, D-37077 G{\"o}t\-tingen, Germany}

\begin{abstract} 

Solar magnetic fields alter scattering polarization in spectral lines like Sr~{\sc i} at 4607\,\AA\ via the Hanle effect, making it a potential diagnostic for small-scale mixed-polarity photospheric magnetic fields.
Recently, observational evidence for scattering polarization in the Sr~{\sc i} 4607\,\AA\ at the solar disk center was found.
Here, we investigate the reliability of the reconstruction method making possible this detection.
To this end, we apply it to linear polarization profiles of the Sr~{\sc i} 4607\,\AA\ line radiation emerging at the disk center obtained from a detailed 3D radiative transfer calculation in a magneto-hydrodynamic simulation snapshot with a small-scale dynamo contribution.
The reconstruction method systematically reduces the scattering amplitudes by up to a factor of two, depending on the noise level. We demonstrate that the decrease can be attributed to two systematic errors: first, the physical constraint that underlies our assumptions regarding the dependence of scattering polarization on the quadrupolar moment of the radiation field, and second, the limitations of our method in accurately determining the sign of the radiation field tensor from the observed intensity image.
However, consistently applying the reconstruction process and after taking into account image degradation effects due to the temporally variable image quality, such as imposed by seeing, observed and synthesized polarization signals show remarkable agreement.
We thus conclude that the observed scattering polarization at solar disk center is consistent with that emerging from magneto-hydrodynamic model of the solar photosphere with an average magnetic field of 170\,G at the visible surface.

\end{abstract}


\section{Introduction} \label{sec:intro}

A major goal in solar physics is to decipher and understand the small-scale magnetism of quiet regions of the solar disk, which covers at least 99\% of the solar lower atmosphere at any given time \citep[e.g., the review by][]{Bellot2019}.
The small-scale magnetic activity in the quiet regions of the solar photosphere could play a significant role in the energy balance of the solar upper atmosphere, as suggested by different investigations \citep[e.g.,][]{TrujilloBueno2004,Rempel2014}.
\citet{Petrovay1993} proposed that small-scale dynamo action is  responsible for the appearance of tangled magnetic fields in the inter-network regions of the solar disk. 
Numerous models of the quiet solar photosphere, based on numerical magneto-hydrodynamic simulations that include small-scale surface dynamo action, exist \citep[e.g.,][]{Vogler2007,Freytag2012}. 
So far, they succeed to statistically reflect the small-scale magnetic topology inferred from observations \citep[e.g.,][]{Danilovic2010a,Bellot2019}. Recently, such computations have been extended to encompass other stars, suggesting that small-scale dynamo action is significant across the lower main sequence, generating small-scale magnetic fields \citep[e.g.,][]{Bhatia2022,Witzke2023}.

However, the simulations are limited to the achievable resolution and, subsequently, the physical scales at which energy is converted and dissipated. 
It is still not fully known if the spatial distributions of magnetic field strengths 
and plasma dynamics seen in small-scale dynamo simulations are compatible with the magnetic fields often operating on spatial scales below the resolution element of a meter-class telescope \citep{Rempel2023}.
Testing the predictions of simulations relies on probing unresolved magnetic fields in the solar atmosphere. 
A suitable observational technique is the study of scattering polarization signals sensitive to the magnetic fields via the Hanle effect \citep{Stenflo1982,TrujilloBueno2004}.

The Sr~{\sc i} line at 4607\,\AA\ shows, among photospheric lines, one of the strongest linear polarization signals when observed close to the solar limb. As a consequence, it has been extensively studied observationally  (\citealt{Stenflo1997a,TrujilloBueno2001,Malherbe2007,Bianda2018,Zeuner2018,Dhara2019}; and \citealt{Zeuner2020}, hereafter Paper I) and theoretically (\citealt{Faurobert-Scholl1993,TrujilloBueno2004, TrujilloBueno2007,DelPinoAleman2018}; and \citealt{DelPinoAleman2021}).

The measurement of the scattering polarization in the Sr~{\sc i} line at 4607\,\AA\ has attracted great interest to probe the spatial structure of unresolved magnetic fields in the photosphere \citep{Bianda2018,Dhara2019,Zeuner2018}. 
Particularly, Paper I recently demonstrated that the scattering polarization signal in the Sr~{\sc i} line is spatially structured on granular scales when observing at the solar disk center. 
\citet{TrujilloBueno2007} theoretically predicted this phenomenon through 3D radiative transfer calculations in a 3D hydrodynamical model of the quiet solar photosphere. More recently, \citet{DelPinoAleman2018} confirmed these findings using a more sophisticated 3D magneto-hydrodynamic model.
The quiet-Sun model by \cite{Rempel2014} used in \cite{DelPinoAleman2018} shows an average magnetic field strength of $\langle B \rangle\approx 170$\,G at the visible surface, produced by small-scale dynamo action and horizontal flux advected in from the lower boundary of the model. 
Combined with the rate of depolarizing collisions with neutral hydrogen atoms for the Sr~{\sc i} line given by \citet{Faurobert-Scholl1995}, the scattering polarization amplitude of the spectral line radiation emerging from this model turns out to be consistent with the center-to-limb variation observed during a minimum and a maximum of the solar cycle \citep[see the information on such observations in Figure 1 of][]{TrujilloBueno2004}. 

A first confrontation of the synthesized scattering amplitudes in the Sr~{\sc i} line at disk center with the observations of Paper I was made by \citet{DelPinoAleman2021}. 
For the first time, the modeling of the Sr~{\sc i} line used for such a comparison were time dependent; therefore, it was possible to take into account the temporal evolution of the solar atmosphere during the observing time. 
One of the main results found by \citet{DelPinoAleman2021} is that the polarization amplitude in the simulation does not significantly evolve during a time period shorter than 5\,min.
However, the observed spatial and spectral characteristics of the polarization signal strongly depend on the observation conditions and setup.

Therefore, for a quantitative comparison with observed scattering amplitudes, it is crucial to accurately incorporate the effects of observation conditions and setup on the spatial and spectral resolution. Degrading spatial dimensions involves considering the resolving power of the telescope and the instrument's sampling, while limiting the spectral resolution includes factors like transmission profiles of filters and the instrument's dispersion power.
A statistical comparison between the degraded and noise-added simulated Stokes $Q/I$ parameter and the observed data suggested that photon noise dominates the observations. However, when comparing a degraded but noise-free simulation with a reconstructed observation\footnote{Using the reconstruction method proposed in Paper I, described later in the paper.} of the scattering polarization, a discrepancy in amplitude of approximately a factor of two was discovered. 
The authors suspected that the source of this systematic error might be related to the reconstruction method \citep[][]{DelPinoAleman2021}. \par

Here, we aim to analyze the reconstruction method employed in Paper I to uncover its potential systematic errors. 
We also compare the (degraded and reconstructed) predicted polarization signals synthesized \citep{DelPinoAleman2018} from the 3D-MHD quiet Sun solar model by \citet{Rempel2014} with observed disk-center reconstructed polarization signals in the Sr~{\sc i} 4607 \AA\ line of Paper I. 
While \citet{DelPinoAleman2021} analyzed a time series, we compare the observations to the single snapshot simulation data of \citet{DelPinoAleman2018}. 
It turns out that the variability of the seeing during the observing time, an effect that we cannot properly take into account when degrading the time series, plays a significant role in the statistical behavior of the resulting polarization signals (see Sect.~\ref{sec:temp}). 
Therefore, there is no significant advantage to carry out the analysis with the time series with respect to the single snapshot. 
Moreover, \citet{DelPinoAleman2021} showed that the temporal evolution minimally affects the polarization amplitudes in observations lasting only a few minutes (as in Paper I). They identified a modest amplitude reduction, not exceeding 15\% relative to the maximum. As we will demonstrate, this reduction is smaller than the systematics inherent in the reconstruction method.

This paper is organized as follows.
In Section~\ref{sec:data_method} we list the main properties of the observed and simulated data, degradation steps applied to the simulation data, and offer a brief explanation of the reconstruction method. Moving on to Section~\ref{sec:results} we implement the reconstruction method on both the simulation and observations and conduct a comparative analysis of their statistical attributes. 

Additionally, we quantify the primary sources of systematic errors, which include the estimation of the radiation field tensors and the temporal variability in the image quality. 
They play a pivotal role in defining the reconstruction method's limitations.
Finally, we summarize our conclusions in Section~\ref{sec:conclusion}.

\section{Data and Methodology} 
\label{sec:data_method}

In this section we present observed data presented in Paper I and simulation data \citep{DelPinoAleman2018} on the intensity and linear polarization of the Sr~{\sc i} line at 4607.3\,\AA. 
We describe the instrumental image degradation effects applied to the simulation to mimic the observations. Finally, we briefly describe the reconstruction method introduced in in Paper I which we apply to both observations and simulation in Sect.~\ref{sec:results}. 

\subsection{Observational and simulation data}
\label{sec:data}
\begin{figure}[ht]
\centering 
 \includegraphics[width=\linewidth]{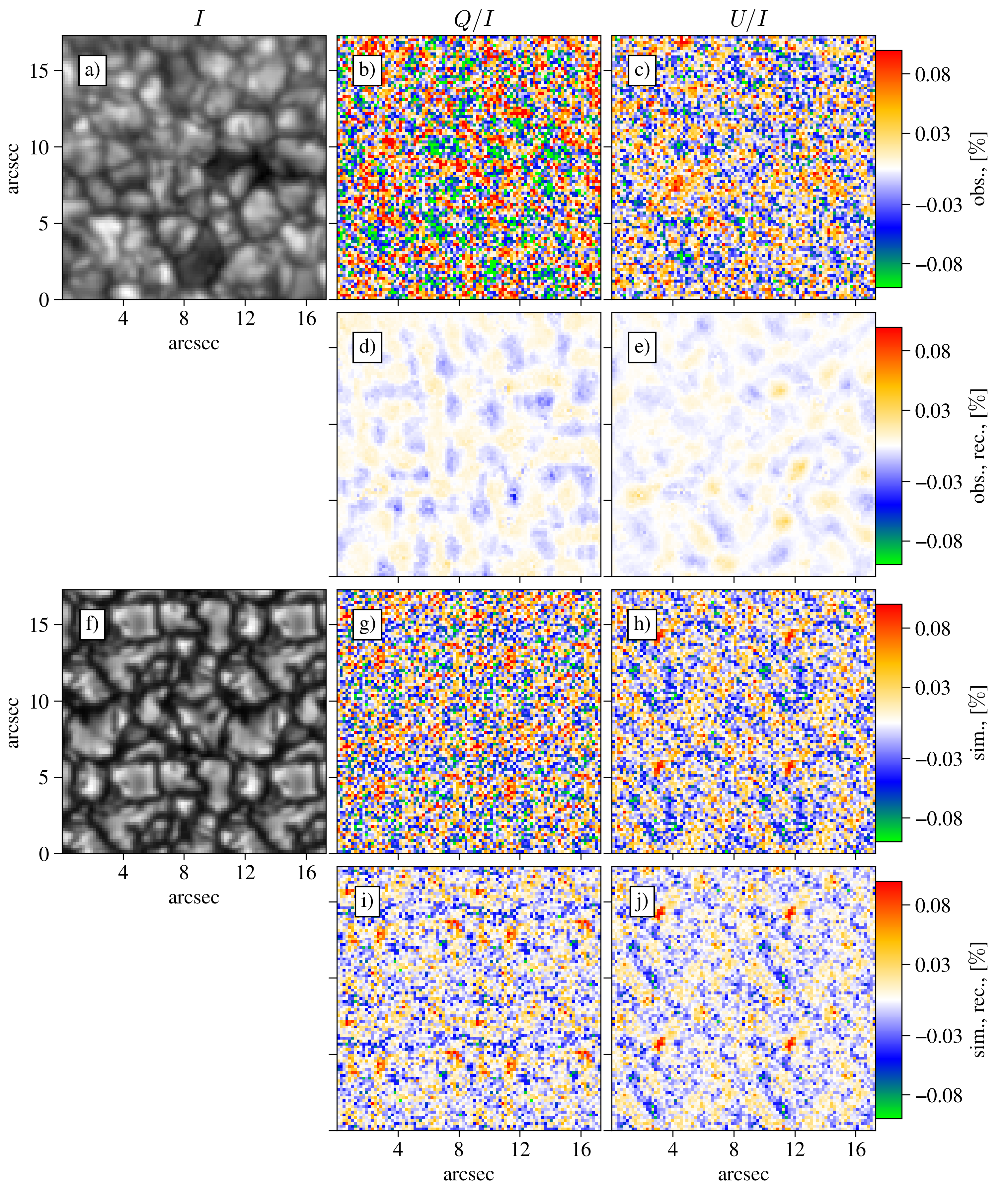}
\caption{Observed and synthetic (degraded to the observation, see text for details) spectrally integrated intensity $I$ and fractional linear polarization $Q/I$ and $U/I$ in the Sr~{\sc i} line around 4607\,\AA.
Observed temporally averaged intensity image (a) and observed Stokes $Q/I$ and $U/I$, b) and c), respectively. d-e) Reconstructed $Q/I$ and $U/I$ spatial maps from the statistical averaged polarization in panels e-f) of Fig.~\ref{fig:j_sum}. The statistical average is based on the $J^2_2$ values (see text for details). These maps have been reconstructed by retracing the accumulated polarization values from Fig.~\ref{fig:j_sum} to their original positions in an observed 5\,s integrated image (i.e., to the places with the corresponding $J^2_2$ values) and then temporally averaged. Degraded single-snapshot theoretical intensity image (f) and Stokes $Q/I$ and $U/I$ images, g) and h), respectively. To facilitate a more straightforward comparison with the observation, the field-of-view is extended periodically. 
i-j) The exact same method as for the observed Stokes images (d-e) is used to reconstruct the spatial distribution of $Q/I$ and $U/I$ from the statistical averaged polarization in panels a-b) of Fig.~\ref{fig:j_sum}.}
\label{fig:newmap}
\end{figure}

The spectropolarimetric data in the Sr~{\sc i} 4607.3\,\AA\ line were obtained using the high-cadence Fast Solar Polarimeter 2 \citep{Iglesias2016,Zeuner2020} attached to the NSF’s Dunn Solar Telescope (DST\footnote{Currently operated by the New Mexico State University.}), located on Sacramento Peak in Sunspot, New Mexico. 
It spans 3.5\,min of observation, with a cadence of 5\,s and 98\% duty cycle.
A single Fabry-P\'{e}rot etalon and a pre-filter were used for the wavelength selection. 
The etalon was tuned very close to the center of the Sr~{\sc i} line at 4607.3\,\AA. 
Due to the broad pre-filter, the spectral resolution was severely degraded.
All the details, especially about the spectral point-spread function (spectral PSF), can be found in Paper I, but we list the most critical parameters for the comparison with the simulation in Table~\ref{tab:obs_simu}.
The (temporally averaged) observed intensity and linear polarization signals $Q/I$ and $U/I$ in the Sr~{\sc i} line are shown in Fig.~\ref{fig:newmap} in the top three panels a), b), and c), respectively. 
In spite of the relatively low noise level, the polarization images lack obvious spatial structures and are noise dominated. 
Hereafter we denote this data set as ``obs.". \par


\citet{DelPinoAleman2018} solved the radiative transfer problem in a 3D radiative magneto-convection simulation calculated with \textit{MURaM} \citep[non-gray radiative MHD code by][]{Vogler2005,Rempel2015}, with a regular spacing of 8\,km with 768 $\times$ 768 $\times$ 137 grid points in the three spatial dimensions \citep{Rempel2014}. 
The vertical dimension was limited to the formation region of the the Sr~{\sc i} line. 
The original MHD model was stored with a time step of about 300~s, but only one snapshot is considered for the spectral line synthesis. 
The main feature of this numerical model is the strong mean magnetic field strength of about 170\,G at the visible surface, resulting from small-scale dynamo action.

The emergent Stokes profiles in the Sr~{\sc i} resonance line were calculated with the two-level module of the PORTA code \citep{Stepan2013}, which solves the non-LTE two-level problem (suitable for this resonance line) taking into account the full 3D structure and dynamics of the model atmosphere, thus accounting for scattering polarization due to the macroscopic velocities and the horizontal inhomogeneities of the model atmosphere\footnote{The public version of the 3D multilevel radiative transfer code PORTA can be found here: \url{https://polmag.gitlab.io/PORTA/}.}. 
The continuum polarization, which at this wavelength is much smaller than that of the Sr~{\sc i} line, is neglected. 
The depolarizing collisional rates are calculated following \citet{Faurobert-Scholl1995}, which result in polarization amplitudes compatible with observed center-to-limb variations.
The full details of the MHD model and of the synthesis of the Sr~{\sc i} 4607.3\,\AA\  can be found in \citet{DelPinoAleman2018} and references therein.

The most important parameters of the synthesized polarization maps are listed in Table~\ref{tab:obs_simu}. 
Several degradation steps are necessary before the synthetic and observed data can be compared. The spatial resolution is lowered with a Gaussian kernel to about 0.4\ensuremath{^{\prime\prime}}. 
The synthetic data is spatially re-sampled by cubic interpolation (and binned) to the observed 0.062\ensuremath{^{\prime\prime}}\,pixel$^{-1}$. 
The spectral sampling is reduced to one point by applying the spectral PSF (with a width of about 67\,m\AA) derived from the observations by convolving the FTS spectrum with spectral transmission profiles of various widths until the observed spectrum was recovered, for details see Paper I. \par
After the previous re-sampling steps, the number of pixels available in the synthetic data is reduced compared to the observational data, due to  the limited size of the simulation box.
This has a clear effect in the statistics when applying the reconstruction method, and we will address it later when describing the method in further detail.
Finally, we add to the synthetic images the same degree of Gaussian noise as present in the observational data.

\begin{deluxetable}{ lcc  }
\tablecaption{Observation and simulation parameters. We assumed 725\,km$\,=\,$1\ensuremath{^{\prime\prime}}. The parameters of the simulated data correspond to the outcomes of the degradation steps (see text). The square brackets enclose} the original properties of the simulation, i.e., before degradation.\label{tab:obs_simu}
\tablehead{
 & \colhead{Observation}& \colhead{Simulation [original]}}

\startdata
  Spatial sampling &  0.062\ensuremath{^{\prime\prime}} pixel$^{-1}$ & 0.062\ensuremath{^{\prime\prime}} pixel$^{-1}$ [8\,km]\\
Resolution &  $\sim$0.4\ensuremath{^{\prime\prime}}& 0.4\ensuremath{^{\prime\prime}} [8\,km]\\
 Spectral sampling   & $\sim$67\,m\AA    & 67\,m\AA \ [3.6~m\AA]\\
 FoV & 17.3\ensuremath{^{\prime\prime}} $\times$ 17.3\ensuremath{^{\prime\prime}}&   8.5\ensuremath{^{\prime\prime}} $\times$ 8.5\ensuremath{^{\prime\prime}} [6144\,km $\times$ 6144\,km]\\
 Temporal sampling & 42 $\times$ 5\,s integration &  single snapshot \\
Noise    &$\sim$0.04\% &  0.04\% [no noise] \\
\enddata
    \end{deluxetable}

We show the synthesized intensity and linear polarization signals after degradation in panels f)-h) of Fig.~\ref{fig:newmap}. Note that the field-of-view (FoV) in this figure is extended only for visualization purposes. 
Hereafter we denote this data set as the simulation or ``sim.". 
We also use a noise-free version of the ``sim." data set, without added (Gaussian) photon noise, which hereafter we denote as ``noisefree" (not shown in the figure). 
The polarization signals are degraded significantly compared to the original simulation, especially when a spectral filter is applied \citep[see][]{DelPinoAleman2018}. 

After degradation, the polarization amplitude is reduced by about a factor of 10. 
Most simulated signals are partially or completely buried within the noise, which is largely consistent with the observed linear polarization images. 
Note that temporal evolution is neglected, and that the intensity image due to the broad spectral PSF in the observation is dominated by the continuum.

\par

To summarize, we have three data sets, each composed of intensity images $I$ and the fractional linear polarization images $Q/I$ and $U/I$ spectrally integrated in the wavelength position of the Sr~{\sc i} line at 4607\,\AA. 

These data sets are the observations and degraded simulation with and without noise. If not specifically stated, we imply that the noisy version of the simulation data set is used. 
The fractional linear polarization from a single snapshot is very similar to a temporally averaged time series of less than 5\,min \citep{DelPinoAleman2021}.
All three data sets are denoted as \textit{originals} on which we will apply the reconstruction method as explained in Sect.~\ref{sec:method}. 

\subsection{Method description} 
\label{sec:method}

In this section we briefly describe the reconstruction method for scattering polarization images in Sr~{\sc i} proposed and described in more detail in Paper I. 
The method's primary goal is to improve the signal-to-noise-ratio for linear polarization. This is achieved by exploiting the deep connection between scattering polarization and the illuminating radiation field, i.e. by spatially averaging pixels that share similar illumination patterns.
The presence of scattering polarization at disk center requires the \emph{axial} symmetry of the radiation field to be broken, and the lowest order of this non-symmetric component can be characterized by a quadrupole. 
The (complex-valued) quadrupole moment of the radiation field is $J^2_2=\tilde{J}^2_2+{\rm i}\hat{J}^2_2$ \citep{LandiDeglInnocenti2004}.
Given an unpolarized incoming radiation, $I(\theta,\chi)$, with $\theta$ and $\chi$ the polar and azimuthal components of the propagation direction,
the complex quadrupolar component of the radiation field is given by an integral over the unit sphere $\mathrm{d}\Omega=\sin\theta \mathrm{d}\theta \mathrm{d}\chi$ \citep{LandiDeglInnocenti2004}\footnote{Note that the different convention of $\mathrm{d}\Omega$ used in Paper I leads to an additional minus sign for $\chi$ in the exponent.}:

	\begin{equation}
	\label{eq:j22}
	J^2_2=\frac{\sqrt{3}}{4}\oint \frac{\mathrm{d}\Omega}{4\pi} \sin^2\theta\, {\rm e}^{2{\rm i}\chi} I(\theta, \chi).
	\end{equation}
 
At disk center, $J^2_2$ is the dominant source of linear polarization, and the theoretical relation $Q/I+{\rm i}U/I \propto -(\tilde{J}^2_2+{\rm i}\hat{J}^2_2)$ applies \citep{DelPinoAleman2018,Zeuner2020}. 
In general, this relation applies only if the structured magnetic field is not dominant. 
The plane of scattering polarization can rotate due to the Hanle effect when a magnetic field exhibits a prominent direction. Recently, the Hanle rotation in the case of Sr~{\sc i} resulting from a magnetic field with a component along the LOS was published for the first time \citep[][]{Zeuner2022}. 
However, the FOV in the observation as well as the simulation exhibit mostly quiet Sun and do not show evident signs of a macroscopic structured magnetic field along the LOS, which was argued in detail in Paper I.

\par

Scattering polarization at disk center is characterized by signatures of positive and negative sign in linear polarization in both Stokes $Q/I$ and $U/I$ parameters \citep{TrujilloBueno2007,DelPinoAleman2018}. 
This excludes the use of conventional spatial averaging techniques, as the risk of signal cancellation is large.
To ensure that the polarization signals add up coherently during averaging, we classify each pixel in the polarization map according to the quadrupole moment of the radiation field illuminating it. 
By having an estimate of $J^2_2$ and especially its sign, 
pixels with similar quadrupole values can be averaged in order to reduce the noise while conserving most of the polarization signal. 
Moreover, by mapping the average linear polarization with the estimated $J^2_2$, we can also approximately reconstruct its spatial distribution.

According to equation~\ref{eq:j22}, $J^2_2$ may be estimated from the Stokes $I$ intensity image. 
In the specific case of the observation obtained in Paper I, Stokes $I$ was dominated by the continuum intensity, attributed to the broad spectral PSF.

There are three steps involved in the reconstruction:
\begin{itemize}

    \item[1)] Estimate the radiation tensor components $\tilde{J}^2_2$ and $\hat{J}^2_2$ from the intensity image. 
    
    \item[2)] Average the fractional linear polarization $Q/I$ and $U/I$ in pixels with similar $\tilde{J}^2_2$ and $\hat{J}^2_2$, resulting in bivariate correlations $Q/I(\hat{J}^2_2,\tilde{J}^2_2)$ and $U/I(\hat{J}^2_2,  \tilde{J}^2_2)$.
    
    \item[3)] Reconstruct the fractional linear polarization by assigning the bin value $Q/I(\hat{J}^2_2,\tilde{J}^2_2)$ and $U/I(\hat{J}^2_2,\tilde{J}^2_2)$ to the pixels with the corresponding $\hat{J}^2_2$ and $\tilde{J}^2_2$ values.
    
\end{itemize}

To estimate $J^2_2$ we consider a thin and uniform scattering layer model at a height $h$ above a series of isotropically radiating atmospheres (i.e., each pixel radiates isotropically the radiation observed at the corresponding LOS). However, the inhomogeneity of the plasma results in a non-isotropic radiation field at each pixel.
For the following analysis, we evaluate equation~\ref{eq:j22} with $h=$ 150\,km to estimate $J^2_2$ at each pixel. 
This value for $h$ is chosen here because it was used in Paper I. 
We further investigate the impact of the parameter $h$ in Section~\ref{sec:reliability_test}. 
$I(\theta, \chi)$ is given by the intensity image (e.g., panels a) and f) of Fig.~\ref{fig:newmap}), for both the observation and the simulation. 
As we are only interested in the relative values of $\tilde{J}^2_2$ and $\hat{J}^2_2$, we normalize them to their respective maximum (in case of the observation, the maximum within the time series). \par

To average the fractional linear polarization $Q/I$ and $U/I$ in pixels with similar $J^2_2$, we create bins with a size of 0.05. 
We then have two bivariate correlations of the fractional linear polarization, $Q/I(\hat{J}^2_2,\tilde{J}^2_2)$ and $U/I(\hat{J}^2_2,\tilde{J}^2_2)$. 
We apply this step to each frame in the observed time series and then average all these bivariate correlations. 
The results are shown in in panels a)-b) and e)-f) of Fig.~\ref{fig:j_sum}, for the simulation and the observation, respectively.  \par
We perform a procedure we call "reconstruction" to estimate the distribution of fractional linear polarization. 
In this process, we reverse the mapping of each $Q/I(\hat{J}^2_2,\tilde{J}^2_2)$ and $U/I(\hat{J}^2_2,\tilde{J}^2_2)$ bin with their corresponding pixel positions. 
As a result, the reconstructed linear polarization maps may contain non-unique pixels. 
We refer to these reconstructed maps as ``rec." 
Examples of reconstructed maps based on simulations can be seen in Fig.~\ref{fig:j_sum} (panels i and j).

\section{Results} 
\label{sec:results}

In this section we study the impact of applying the reconstruction method outlined in Sect.~\ref{sec:method} to the data presented in Sect.~\ref{sec:data}. 
This exploration enables us to not only quantitatively compare the observations to the simulations but also to assess the reliability of the reconstruction process. 
Furthermore, having access to the radiation field tensor components $\tilde{J}^2_2$ and $\hat{J}^2_2$ in the simulation provides an opportunity to evaluate the accuracy of estimating these tensors, a crucial aspect of the reconstruction method.
We focus our analysis on four topics:

\begin{itemize}

    \item[1.] The quantitative comparison between simulated and observed bivariate correlations $Q/I(\hat{J}^2_2,\tilde{J}^2_2)$ and $U/I(\hat{J}^2_2,\tilde{J}^2_2)$.
    
    \item[2.] Estimating the impact of temporal image degradation on the observed fractional linear polarization amplitude.
    
    \item[3.] The comparison of the reconstructed and original linear polarization spatial distributions in the simulation, to determine the accuracy of the reconstruction.
    
    \item[4.] The comparison of the $\tilde{J}^2_2$ and $\hat{J}^2_2$ estimation to the actual values given by the simulation, to assess the estimation accuracy.
    
\end{itemize}

\subsection{Quantitative comparison between simulation and observation}
\label{sec:comp}

\begin{figure}
\centering 

\includegraphics[width=\linewidth]{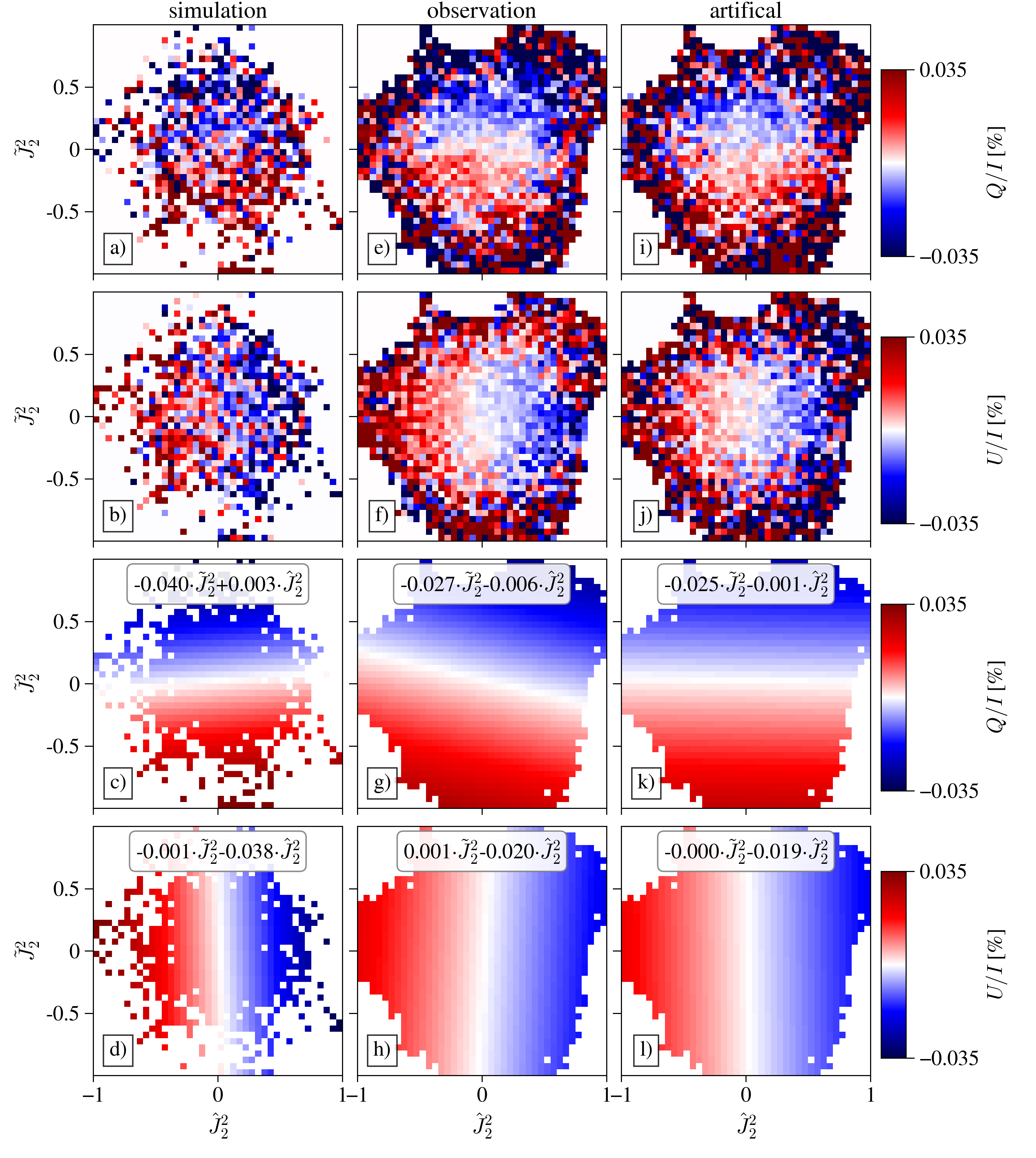}
    \caption{
    Bivariate correlation of the linear polarization $Q/I$ (first row) and $U/I$ (second row) with the $J^2_2$ radiation field tensor. The third (fourth) row shows a two-dimensional plane fit to the panels of the first (second) row. From left to right, we show the bivariate correlation for the simulation (left column, constructed from panels f)-h) of Fig.~\ref{fig:newmap}), the observation (center column, constructed from panels a)-c) of Fig.~\ref{fig:newmap}), and for an artificial observation (right column, see Sect.~\ref{sec:temp} for further details).
    } 
\label{fig:j_sum}
\end{figure}

We show the bivariate correlations $Q/I(\hat{J}^2_2,\tilde{J}^2_2)$ and $U/I(\hat{J}^2_2,\tilde{J}^2_2)$ in Fig.~\ref{fig:j_sum} for both the simulation (panels a)-b)) and the observations (panels e)-f)). 
The quadrupolar components $\tilde{J}^2_2$ and $\hat{J}^2_2$ of the radiation field are normalized to their maximum value. 
As a result, the bivariate correlations exhibit a consistent distribution, regardless of a specific granulation pattern.
 
Note that the total duration of the observation is sufficiently short as to conserve the same granulation pattern.
Therefore, it is possible to compare the simulation and observation quantitatively, without any bias due to the particular realization of the solar photosphere they correspond to. 
The main noticeable difference between the left ``sim." and ``obs." columns in Fig.~\ref{fig:j_sum} is the $\hat{J}^2_2-\tilde{J}^2_2$ plane coverage, which is larger for the observation due to the significantly larger number of available pixels in the whole time series.
\par

We find a strong anti-correlation between the fractional linear polarization signals and the radiation field quadrupolar components, both in the observation and in the simulation ($Q/I$ is anti-correlated with $\tilde{J}^2_2$, while $U/I$ is anti-correlated with $\hat{J}^2_2$).  
The orientation of the red-blue pattern in Fig.~\ref{fig:j_sum} is almost perfectly aligned with the $\hat{J}^2_2$ and $\tilde{J}^2_2$ axes for both observations and simulations. 
This indicates that no organized and large scale mean magnetic field is present in either the observations or the simulations, as this would rotate the red-blue pattern in the $\hat{J}^2_2-\tilde{J}^2_2$ plane via the Hanle effect. The small rotation seen in the observation (of about 4.5$^\circ$) is within observational uncertainties (mainly due to the statistical noise, note that $Q/I$ has a larger noise level than $U/I$). 
\par

We fit a weighted two-dimensional surface to the bivariate correlations of the fractional linear polarization in Fig.~\ref{fig:j_sum} to quantify the differences between the observed and the simulated signals. 
We show the fitted surfaces in the bottom six panels of Fig.~\ref{fig:j_sum} (with the coefficients of the fits in the legend of each panel).
The coefficient of this fit is two-times larger in the simulation $U/I(\hat{J}^2_2,\tilde{J}^2_2)$ than in the observation $U/I(\hat{J}^2_2,\tilde{J}^2_2)$. 
Regarding $Q/I(\hat{J}^2_2,\tilde{J}^2_2)$, they differ by about 30\%, being larger in the simulation.

Solar temporal evolution can be excluded as the origin for this discrepancy according to the analysis of \citet{DelPinoAleman2021}, which can only account for a maximum of 15\% polarization amplitude reduction. 
One possible reason could be the variation of the image quality in the observation during the observation time due to the seeing. 
Another potential reason could be a larger strength of the unresolved magnetic field in the observed region of the photosphere relative to that in the magneto-hydrodynamic simulation. 
We further investigate these possibilities in Sect.~\ref{sec:temp}. \par

Finally, we reconstruct the linear polarization spatial distribution as described in Sect.~\ref{sec:method} using the data shown in Fig.~\ref{fig:j_sum}. 
For both the simulation and observation, this reconstruction involves mapping averaged $Q/I(\hat{J}^2_2,\tilde{J}^2_2)$ and $U/I(\hat{J}^2_2,\tilde{J}^2_2)$ values to their corresponding positions in the original images.
The resulting polarization maps are shown in Fig.~\ref{fig:newmap}. 
These reconstructed maps successfully reveal previously obscured polarization features that were hidden within the noise. 
We further assess the reliability of the reconstruction in the simulation by comparing it to the noise-free data later in the paper.
\par

\begin{deluxetable}{ cc c c|  c c c   }
\tablecaption{Absolute mean and root-mean-square (RMS) values of the fractional linear polarization images before  and after reconstruction for the observations and simulation. 
\label{tab:statistics}}
\tablehead{
  &\multicolumn{3}{c}{before reconstruction} & \multicolumn{3}{c}{after reconstruction}\\
                  \colhead{}  &\colhead{obs.}  & \colhead{sim., noisefree} & \colhead{sim.} & \colhead{obs., rec} & \colhead{sim., noisefree, rec.} & \colhead{sim., rec.} }
\startdata
 mean, $|Q/I|$  [\%]  & 0.03 &  0.014 & 0.037 & 0.005 & 0.010 & 0.014 \\ 
 RMS,  $Q/I$    [\%]  & 0.06 &  0.018 & 0.046 & 0.006 & 0.014 & 0.020 \\ 
 mean, $|U/I|$  [\%]  & 0.05 &  0.014 & 0.027 & 0.003 & 0.010 & 0.011 \\
 RMS,  $U/I$    [\%]  & 0.04 &  0.018 & 0.034 & 0.005 & 0.013 & 0.016 \\
\enddata
    \end{deluxetable}

In general, the spatial structures of the reconstructed polarization show very similar sizes and distributions in the simulation and observation. 
To quantify the spatial distribution, we use the Relevance Vector Machine implementation\footnote{\url{https://github.com/aasensio/rvm}} briefly explained in \citet{Campbell2021} to analyze the reconstructed polarization maps. 
This code has the primary goal to identify the most relevant vectors from a any basis. 
Since the reconstructed polarization maps appear to be very periodic, our basis consist of $\sin(xf)$ and $\cos(xf)$ functions of different frequencies $f$, with $x$ being the spatial coordinate. 
The most relevant period for the observation is 0.4\,arcsec, while for the simulation it is 0.43\,arcsec. 
This is consistent with the polarization pattern seen with the naked eye in the reconstructed maps.
The absolute mean and RMS values of all original and reconstructed polarization images are compiled in Table~\ref{tab:statistics}. 
The observed reconstructed polarization images exhibit lower contrast compared to the simulated ones.
To elaborate, the mean total linear polarization ($L=\sqrt{Q/I^2+U/I^2}$) for the observation is 0.004\% after reconstruction, while for the simulation, it amounts to 0.018\%. 
This implies that the reconstructed simulation displays polarization amplitudes more than four times greater than those observed. 
This discrepancy cannot be attributed to the reconstruction method, as it is consistently applied to both the simulation and observations.
Furthermore, we conducted tests by using the time series of the simulation from \citet{DelPinoAleman2021}, and as expected, the reduction of the polarization amplitude was negligible.
However, the seeing plays a crucial role in the observed linear polarization amplitude \citet{DelPinoAleman2018}. 
In the next section we therefore investigate if the image degradation in the observation due to seeing can account for this difference.

\subsubsection{Temporally variable image degradation}
\label{sec:temp}

Since the reconstruction relies on the bivariate correlations in Fig.~\ref{fig:j_sum}, we can investigate the disparities between the fitted coefficients to understand the differences between the observation and the simulation.
We suspect that the variability in the observations, wherein each image in the time series exhibits slightly different $\tilde{J}^2_2$ and $\hat{J}^2_2$ distributions, contributes significantly to the weakening of the coefficients.
It is worth noting that not all potential $\tilde{J}^2_2$ and $\hat{J}^2_2$ values are simultaneously present in all images of the time series.

$\hat{J}^2_2$ and $\tilde{J}^2_2$ are normally distributed, meaning that larger $\hat{J}^2_2$ and $\tilde{J}^2_2$ values are less common. During the averaging process, fewer pixels contribute to the calculation at large $\hat{J}^2_2$ and $\tilde{J}^2_2$ values. 
As a result, statistical noise is more pronounced at larger $\hat{J}^2_2$ and $\tilde{J}^2_2$ values, resulting in a noisy ``ring'' evident in panels e)-f) of Fig.~\ref{fig:j_sum}.
This noisy ring decreases the steepness of the slope in the fit and, as a result, the fitted coefficients are smaller. 
To take into account this effect and to test if the coefficients of the single snapshot simulation are compatible with the observations, we generate artificial bivariate correlations $Q/I(\hat{J}^2_2,\tilde{J}^2_2)$ and $U/I(\hat{J}^2_2,\tilde{J}^2_2)$.  \par

To generate the artificial observation we {\em assume} that, for each ${J}$ image in the observation, the polarization pattern is given by the coefficients from the fitted surfaces of the bivariate correlations of the simulation. 
The ``artificial'' fractional linear polarization images with scattering signals $(Q/I)_{\mathrm{art.}}$ and $(U/I)_{\mathrm{art.}}$ are therefore generated by:
\begin{equation}
\label{eq:1}
\left(Q/I (x,y,t)\right)_{\mathrm{art.}} =\alpha_{Q,\mathrm{sim.}}\cdot\left(\tilde{J}^2_2(x,y,t)\right)_{\mathrm{obs.}} +\beta_{Q,\mathrm{sim.}}\left(\hat{J}^2_2(x,y,t)\right)_{\mathrm{obs.}} + \mathrm{noise}
\end{equation}
and 
\begin{equation}
\label{eq:2}
\left(U/I (x,y,t)\right)_{\mathrm{art.}} = \alpha_{U,\mathrm{sim.}}\cdot\left(\tilde{J}^2_2(x,y,t)\right)_{\mathrm{obs.}} +\beta_{U,\mathrm{sim.}}\left(\hat{J}^2_2(x,y,t)\right)_{\mathrm{obs.}}+ \mathrm{noise}
\end{equation}

for each time step $t$ and each spatial pixel $(x,y)$. The coefficients of the simulation are $\alpha_{Q,\mathrm{sim.}}$=0.039, $\beta_{Q,\mathrm{sim.}}$=0.001, $\alpha_{U,\mathrm{sim.}}$=0.0, and $\beta_{U,sim.}$=0.041 (see Fig.~\ref{fig:j_sum}). Of course, the statistical photon noise has to be added.

Subsequently, we employ the same averaging procedure concerning $\hat{J}^2_2$ and $\tilde{J}^2_2$, mirroring the approach used for the observation. 
The outcomes are illustrated in panels i)-l) of Fig.~\ref{fig:j_sum}.

This approach provides a significantly improved representation of the temporal degradation effects in the observed data, surpassing what can be accomplished by degrading a simulated time series. 
In the context of the real observations, various factors like atmospheric turbulence and static image degradation introduce complexities that make it challenging to create a faithful degraded simulation. These issues often necessitate statistical modeling, but the statistics underlying these real-world effects are not always well-known. 
Consequently, modeling the observation typically involves numerous additional assumptions and comes with inherent uncertainties.
In contrast, our method for generating artificial observations is straightforward. It hinges on a single assumption: that the polarization characteristics in each individual image within the observational time series exhibit a dependence on $J^2_2$ akin to the behavior found in the simulation. 
This simplifies the modeling process, eliminates the need for elaborate statistical modeling, and offers a more direct path to capturing temporal image degradation.

The most remarkable result is that the coefficients are reduced by almost a factor of two after averaging the 42 realizations (i.e., from $\beta_{U,sim.}$=0.041 to $\beta_{U,art.}$=0.021, see Fig.~\ref{fig:j_sum}). 
The obtained fit coefficients are remarkably close to those of the observation. 
They differ by less than 10\%. 
This result shows that the significant impact of image degradation effects on the observed scattering polarization amplitudes.
Hence, for meaningful comparisons between two data sets, it is crucial to not only apply spectral and spatial degradation, but also to consider for possible temporal image degradation effects.

However, this experiment also demonstrates that the polarization amplitudes in the observations are compatible with the simulation once all degradation effects are taken into account. 
We emphasize that the artificial observation does not contain any critical information on the polarization signal from the observations. 

\subsection{Reliability of the reconstruction method}
\label{sec:reliability_test}

To evaluate the quality of the reconstruction we statistically compare the simulated noise-free polarization images with their reconstructions.
Secondly, we check which structures are more reliably reconstructed.\par

\begin{figure}
\gridline{\fig{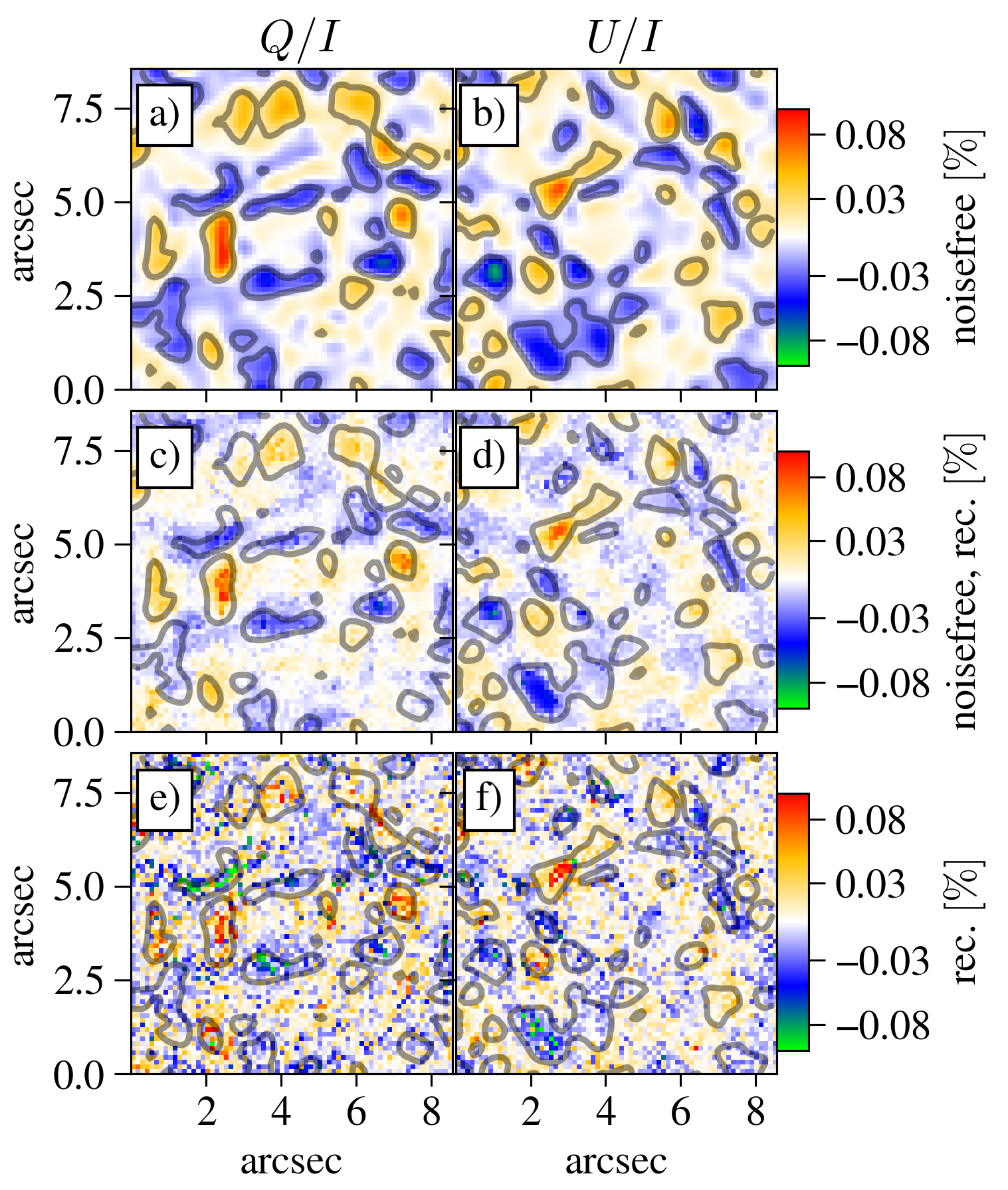}{0.5\textwidth}{(i)}
          \fig{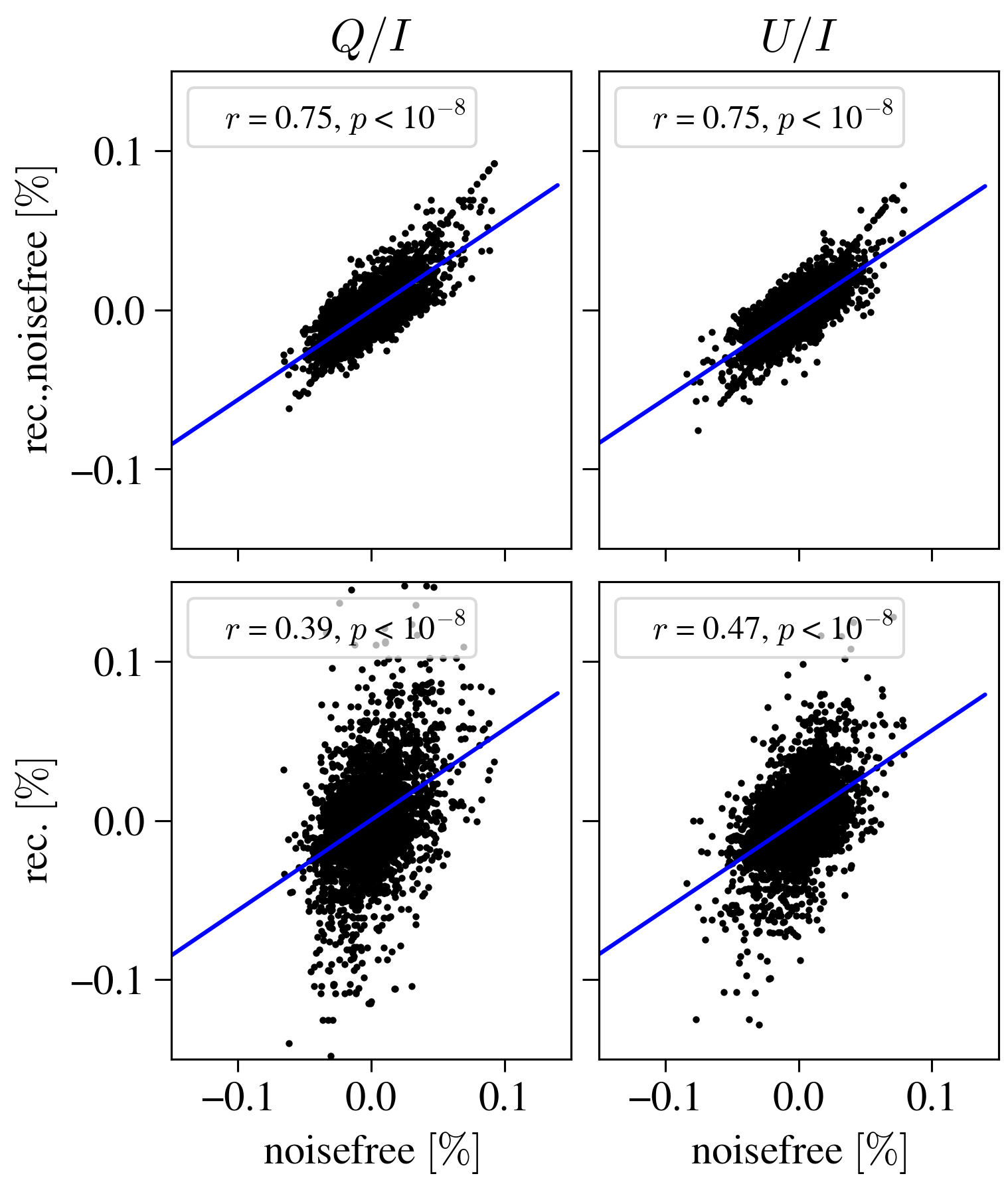}{0.5\textwidth}{(ii)}}
\caption{Left panels (i): Original and reconstructed simulated linear polarization $Q/I$ (left column) and $U/I$ (right column) images. a)-b) Noise-free data. c)-d) Reconstructed polarization maps from the noise-free images. e)-f) Reconstructed polarization maps from the images containing noise. 
The grey contours indicate a polarization amplitude of 0.02\%. 
Right panels (ii): Scatter plots of the reconstructed $Q/I$ (left column) and $U/I$ (right column) amplitudes without (top row) and with (bottom row) noise, against the corresponding polarization amplitude in the original simulation snapshot without noise (but degraded). 
Only pixels inside the grey contours in panels (i) are considered. 
The blue lines show the linear fit, the correlation coefficients $r$ and the corresponding $p$ values are given in the inset.}
  \label{fig:countour_bar}
\end{figure}

When comparing the simulated noise-free polarization images in panels a)-b) of Fig.~\ref{fig:countour_bar} with their respective reconstructed images in panels c)-d) of Fig.~\ref{fig:countour_bar}, we find that the reconstructed polarization patterns closely match the original ones, especially when the polarization patches have a size of $1\ensuremath{^{\prime\prime}} - 2\ensuremath{^{\prime\prime}}$ with amplitudes of 0.02\% and larger. 
However, small-scale details get lost during the reconstruction process. 
Therefore, we find that the reconstruction works more reliable for larger spatial areas with  stronger polarization signals, as could be expected.

It is noteworthy that the reconstruction based on the noisy polarization image in panels e)-f) of Fig.~\ref{fig:countour_bar} seems very similar to the reconstruction of the noise-free images in panels c)-d) of Fig.~\ref{fig:countour_bar}. 
This indicates that the reconstruction is robust for the noise level of the observation.

\par

In quantitative terms, the reconstructed images in Fig.~\ref{fig:countour_bar} show less contrast than the original noise-free polarization image. 
In fact, the mean value of the polarization amplitudes is reduced by about 30\% when reconstructed (see noise-free values in Table~\ref{tab:statistics}).  
This reduction is supported by the scatter plots shown in Fig.~\ref{fig:countour_bar}.
Because the reconstruction shuffles the polarization values based on the estimated radiation field tensor $J^2_2$, originally weaker polarized regions may contain more polarization after reconstruction and vice-versa. 
This effect is amplified by the noise, therefore weaker polarized regions show more polarization than expected, see the bottom scatter plot in Fig.~\ref{fig:countour_bar}.

Additionally, there is likely also some amplitude reduction because of the miss-identification of pixels (see Sec.~\ref{sec:j22}). 
For example, additional polarizing mechanisms, such as spatial gradients of the horizontal plasma velocities, which are not taken into account by the reconstruction model, result in such a miss-identification of pixels. 
However, also wrong assignments of the sign of a pixel play a significant role during the reconstruction process.
Remarkably, only very few pixels change their sign during the reconstruction process,
about 6\% of the pixels for the noise-free case and about 20\% for the noisy case. 
In the following, we investigate the source for these errors by examining our assumptions for the reconstruction method, i.e. what is the error in our estimation of $J^2_2$ and how well is the linear polarization actually correlated with it.

\subsubsection{Comparison of simulated and estimated $\hat{J}^2_2$ and $\tilde{J}^2_2$ from the radiative transfer solution} 
\label{sec:j22}

\begin{figure}[ht]
\centering 
  \includegraphics[width=\linewidth]{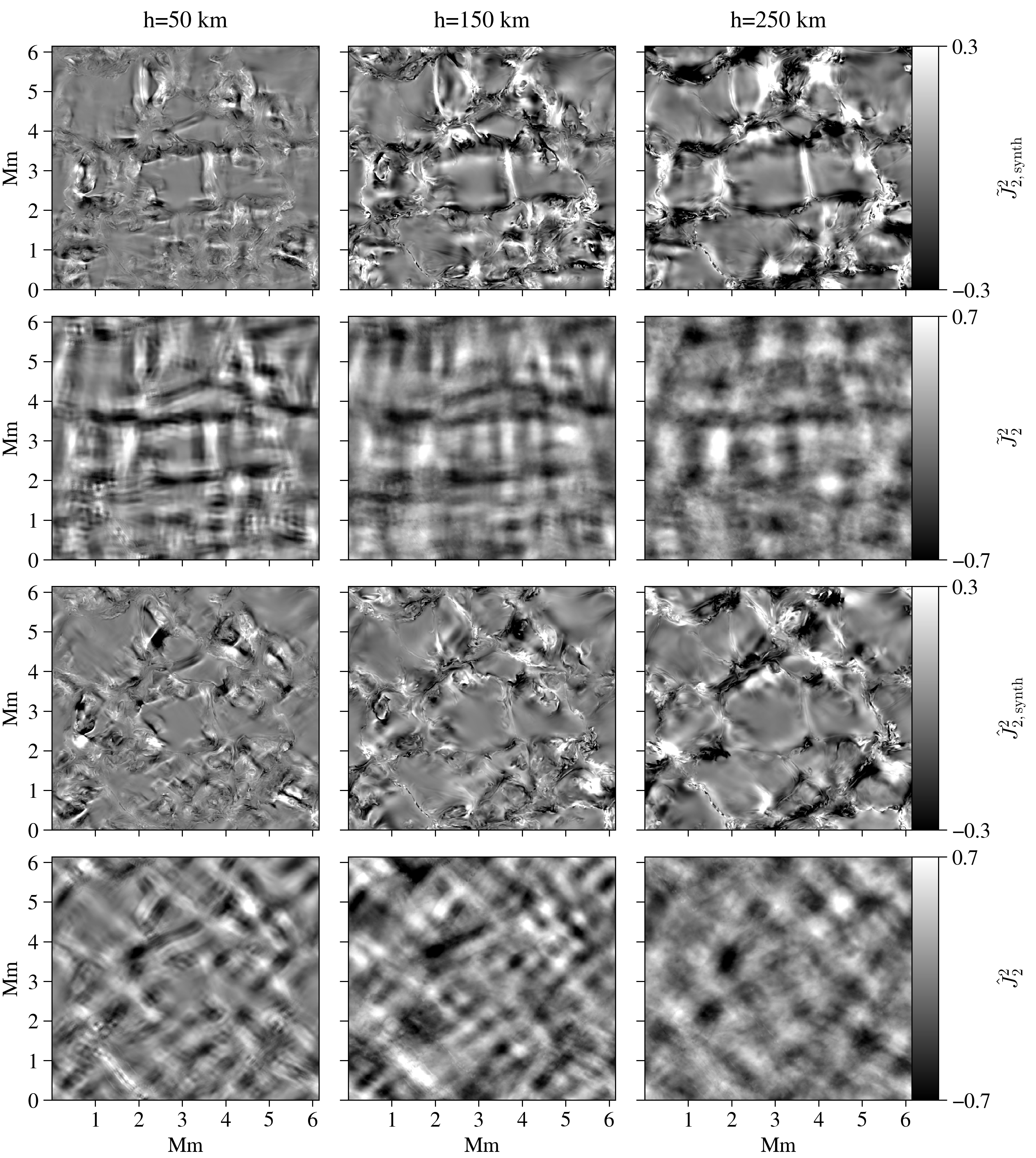}
\caption{Real and imaginary parts of the radiation field tensor component $J_2^2$ as given by the simulation (first and third rows) and estimated from just the intensity image (second and fourth rows) at different heights, from left to right: 50, 150, and 250\,km. All images are normalized to their maximum value.}
\label{fig:j22_est_orig}
\end{figure}

To estimate the source of the sign-flip error, we analyse and compare the $\hat{J}^2_2$ and $\tilde{J}^2_2$ maps from the simulation with our estimation as described in Sect.~\ref{sec:method}. 

We carry out this comparison with the estimated radiation field tensor components at three different geometric heights $h$. 
The estimated $\hat{J}^2_2$ and $\tilde{J}^2_2$ values are calculated with equation \ref{eq:j22}, while the simulated $\hat{J}^2_{2,\mathrm{synth}}$ and $\tilde{J}^2_{2,\mathrm{synth}}$ values at disk center ($\mu=1$) are given by \citep[from the Appendix of][]{DelPinoAleman2018}:

\begin{subequations}
\label{eq:j22_sim}
\begin{align}
\hat{J}^2_{2,\mathrm{synth}}  & = \frac{\sqrt{3}}{4}\int \mathrm{d} \nu \oint \phi (\nu,\Omega) \frac{\mathrm{d}\Omega}{4\pi} \Big[ \cos{(2\chi)} \left[\sin^2{\theta} I(\nu,\Omega) - (1 + cos^2{\theta}) Q(\nu,\Omega) \right] + 2 \sin{(2\chi)}\cos{\theta} U(\nu,\Omega)  \Big], \\
\tilde{J}^2_{2,\mathrm{synth}} & = \frac{\sqrt{3}}{4}\int \mathrm{d} \nu \oint \phi (\nu,\Omega) \frac{\mathrm{d}\Omega}{4\pi} \Big[ \sin{(2\chi)} \left[\sin^2{\theta} I(\nu, \Omega) - (1 + \cos^2{\theta}) Q(\nu, \Omega)\right] - 2 \cos{(2\chi)}\cos{\theta}U(\nu, \Omega)\Big],
\end{align}
\end{subequations}

where $\phi$ is the Voigt absorption profile, with the frequency $\nu$ and the propagation direction of the radiation $\Omega$ characterized by the polar and azimuthal angles $\theta$ and $\chi$, respectively. 
Note that we have omitted the spatial dependence ($x,y,z$) of $\hat{J}^2_{2,\mathrm{synth}}$, $\tilde{J}^2_{2,\mathrm{synth}}$, $\phi$, $I$, $Q$, and $U$ in these equations.
We extracted $\hat{J}^2_{2,\mathrm{synth}}$, $\tilde{J}^2_{2,\mathrm{synth}}$ at three different heights: 50\,km, 150\,km and 250\,km.
\par

We keep the original spatial resolution of 8\,km for the $\hat{J}^2_{2,\mathrm{synth}}$ and $\tilde{J}^2_{2,\mathrm{synth}}$ from the simulation, as well as for the intensity image from which we estimate $\hat{J}^2_2$ and $\tilde{J}^2_2$, to assess the conservation of details in the estimation process.
We estimate $\hat{J}^2_2$ and $\tilde{J}^2_2$ at three heights $h$, effectively estimating them in proximity to the heights where ${J}^2_{2,\mathrm{synth}}$ was extracted from the simulation.

The simulated and estimated radiation field tensors are shown in Fig.~\ref{fig:j22_est_orig}. 
Note that \citet{DelPinoAleman2018} extracted $\hat{J}^2_2$ and $\tilde{J}^2_2$ at the corrugated surface with $\tau=1$ at the line center, therefore a few differences may be expected and noticed when compared to the $\hat{J}^2_{2,\mathrm{synth}}$ and $\tilde{J}^2_{2,\mathrm{synth}}$ images presented here. 

The $\hat{J}^2_{2,\mathrm{synth}}$ and $\tilde{J}^2_{2,\mathrm{synth}}$ from the simulation presented here are integrated over the absorption profile.   \par

The most important aspect to verify is how well the signs of $\hat{J}^2_2$ and $\tilde{J}^2_2$ are estimated by Eq.~\ref{eq:j22}, because this effectively determines the degree of cancellation during the averaging process. 
The fractional number of pixels with different signs are given in Table~\ref{tab:sign_error} (averaged for both signs as well as the real and imaginary part) for each of the three heights. 

\begin{table}[h!]
\centering
\begin{tabular}{ c|c|c  }
height  & sign(${J}^2_2$) $\neq$ sign(${J}^2_{2,\mathrm{synth}}$) & sign($\{Q,U\}_{\mathrm{simu,noisefree}}$) $\neq$  sign($-{J}^2_{2,\mathrm{synth}}$)\\
 \hline
  50\,km & 16\% & 32\% \\
 150\,km & 23\% & 24\% \\
 250\,km & 26\% & 15\% \\
\end{tabular}
 \caption{Sign error rates for ${J}^2_2$ and simulated fractional linear polarization, where the reference sign is given by ${J}^2_{2,\mathrm{synth}}$ at each height. The rates are averaged for both signs and either real and imaginary parts in the case of ${J}^2_2$, or both linear polarization states. }
 \label{tab:sign_error}
\end{table}

The smallest error in the $J^2_2$ sign is found for the smallest height (the deepest layer). 
This is expected because as we delve deeper into the atmosphere, the radiation field tends to exhibit more isotropic characteristics per resolution element.

Consequently, the assumption that each pixel radiates isotropically aligns more closely with reality in the deeper layers of the atmosphere. However, each pixel radiates differently, resulting in an overall non-isotropic radiation field.
Moreover, the farther from the surface, the greater the expected impact of radiation transfer effects. These effects are anticipated not only to modify the continuum radiation but also to radiatively couple the regions in the atmosphere above the surface.
Remarkably, this result demonstrates that the $\hat{J}^2_2$ and $\tilde{J}^2_2$ estimation correctly identifies the sign for about 75\% or more of the pixels, independently of the chosen geometrical height. \par

However, one additional critical assumption influencing the result of the averaging process is that $Q/I \propto - \tilde{J}^2_2$ and $U/I \propto - \hat{J}^2_2$, which means that the linear polarization parameters have the opposite sign to the components of ${J}^2_2$. 

The resulting sign error rates from comparing the linear polarization states with $\hat{J}^2_{2,\mathrm{synth}}$ and $\tilde{J}^2_{2,\mathrm{synth}}$ are shown in Table~\ref{tab:sign_error}. 
For $Q/I$ ($U/I$), the majority of pixels show the opposite sign in the $\hat{J}^2_{2,\mathrm{synth}}$ ($\tilde{J}^2_{2,\mathrm{synth}}$) image, because the error rates are well below 50\%. 
Therefore, we find that this is the fundamental limitation of the reconstruction method: even if there were a perfect match between the estimated and ``true'' radiation tensor component signs, 
a difference of approximately 15-32\% in the sign of the scattering polarization with respect to radiation field tensor components  persists (according to this simulation). 
The difference decreases towards the higher layer of 250\,km. 
This means that other sources, sinks, or alterations of scattering polarization \citep[like velocity fields or magnetic fields, see][]{DelPinoAleman2018} have a significant contribution to the scattering polarization amplitude. 
For the height of 150\,km, which we used for the analysis in the sections before, the sign mismatch affects, on average, the 24\% of all pixels under consideration. 
A sign mismatch of 24\% means a maximum amplitude decrease of approximately 48\% during averaging, because each pixel which has the opposite sign cancels another pixel with the expected sign. On average, taking into account both sign errors, the amplitude reduction is 47\% when considering this height. 
However, the final polarization amplitude reduction depends on the ratio of strong signals to weak signals within the FoV, and the noise level (see discussion above and Fig.~\ref{fig:countour_bar}).

\section{Discussion and conclusion} 
\label{sec:conclusion}

We carefully compared the scattering polarization patterns predicted by the simulation, using synthetic Sr~{\sc i} line 4607\,\AA \ fractional linear polarization spectra at disk center calculated by \citet{DelPinoAleman2018}, and the observed polarization signals obtained by \citet{Zeuner2020}. 
All pertinent instrumental image degradation effects were taken into consideration.
Additionally, we evaluated the performance of the technique described in \citet{Zeuner2020} to reconstruct fractional linear polarization images at the center of the Sr~{\sc i} for a disk center line-of-sight.

Upon comparing the reconstruction (employing the reconstruction method consistently) of both observations and simulation, we made the following observations:

\begin{itemize}

    \item The spatial distribution of linear polarization in the observations shows a lattice-like pattern very similar to that predicted by the simulation, both with a characteristic spatial scale of about 0.4\,arcsec.
    
    \item After taking into account all instrumental image degradation effects, the polarization amplitude of the simulation is three times larger than that in the observation. We identified the most likely cause to be the time variation of the image quality (e.g., seeing) during observations.
    
    \item After additionally accounting for image degradation due to time-variable seeing, we discovered that the polarization amplitude in the simulation aligns closely (within a 5\% margin) with the observations.
    
\end{itemize}

The first finding is consistent with the theoretical investigations of the fractional linear polarization in the Sr~{\sc i} line at disk center by \citet{TrujilloBueno2007, DelPinoAleman2018}; and \citet{DelPinoAleman2021}. 
Based on the last finding, the high level of consistency between the simulated and observed fractional polarization amplitudes provides valuable evidence that the small-scale magnetic field characteristics in the observed quiet Sun region statistically resemble those present in magneto-convection simulations, like those performed by \citet{Rempel2014}. Notably, in these simulations, a portion of the magnetic field is generated through small-scale dynamo action.

This supports the suggestion that small-scale dynamo action could be a significant source of the magnetic field in the quietest regions of the Sun \citep{Emonet2001,SanchezAlmeida2003,TrujilloBueno2004,Lites2014}. 
The magneto-convection model on which this investigation is based has a mean average surface magnetic field of $\langle B \rangle\approx 170$\,G. 
\citet{DelPinoAleman2018} demonstrated that the Hanle depolarization produced by the small-scale magnetic field of this 3D model is such that the scattering polarization amplitudes of the Sr~{\sc i} 4607 \AA\ line observed without spatio-temporal resolution can be fitted when using the elastic collisional rates obtained by \citet{Faurobert-Scholl1995} (see their figure 12), but \citet{DelPinoAleman2018} also showed that slightly larger magnetic field strengths would be needed if the elastic collisional rates proposed by other authors were more accurate (see their figure 13). 
It would be of interest to carry out investigations similar to the present one, but using other MHD models with different magnetic field strengths and distributions.
In any case, we believe that the conclusion of \citet{TrujilloBueno2004} that the inter-network regions of the quiet Sun carry a substantial amount of ``hidden"  magnetic energy due to the presence of an unresolved small-scale magnetic field with $\langle B \rangle$ of the order of 100\,G has been confirmed by \citet{DelPinoAleman2018} and by the present paper.    

\par

We evaluated the performance of the reconstruction method by comparing the direct output of the simulated fractional polarization with its reconstruction. We found the following:

\begin{itemize}

    \item The spatial distribution of scattering polarization is, in general, well reconstructed.
    
    \item The noise level of the observations by \citet{Zeuner2020} has only a minor impact on the reconstruction, i.e., decreasing the contrast.
      
    \item Larger ($\sim1\ensuremath{^{\prime\prime}}$) and stronger polarimetric structures tend to be more reliably reconstructed than weak and smaller structures due to cancellation effects.
    
   \item The reconstruction introduces a systematic error which decreases the mean absolute linear polarization amplitude by more than 40\%.
   
\end{itemize}

The first and second findings give us confidence that the maps reconstructed from observations in \citet{Zeuner2020} correctly reproduce the structures buried within the noise.  Of course, the quality of the reconstruction still depends on the noise level of the original data.
To understand the source of the described systematic error, we examined the radiation tensor elements $\hat{J}^2_2$ and $\tilde{J}^2_2$. Specifically, we compared the estimated signs of $\hat{J}^2_2$ and $\tilde{J}^2_2$, both based on the simulated intensity image, with the actual values of $\hat{J}^2_{2, \mathrm{synth}}$ and $\tilde{J}^2_{2, \mathrm{synth}}$ derived during the radiation transfer problem solution.
We also compared the $\hat{J}^2_{2,\mathrm{synth}}$ and $\tilde{J}^2_{2,\mathrm{synth}}$ with the simulated linear polarization images $Q/I$ and $U/I$. 
This comparison holds significant importance because we base our assumption for the reconstruction method on the idea that the linear polarization Stokes parameters are proportional to these radiation field tensor components. Our findings are as follows:

\begin{itemize}

  \item The sign of the estimated $\hat{J}^2_2$ and $\tilde{J}^2_2$ differs from those in the simulation for less than 25\% of the pixels at the height considered in \citet{Zeuner2020}. The error increases with height.
 
  \item The sign of the linear polarization, when compared to the $-\hat{J}^2_{2, \mathrm{synth}}$ and $-\tilde{J}^2_{2, \mathrm{synth}}$ tensor components, differs also in less than 25\% of the pixels. The error decreases with height.
  
\end{itemize}

The combination of the two sign errors above are the main contributions to the systematic error and may lead to the conclusion in \citet{Zeuner2020} that the reconstruction gives the best results for an assumed scattering layer height of 150\,km. This could be indicative of a compromise between the two opposite behaviors of the sign of actual scattering polarization signals and the estimated radiation field tensors, as the estimation is better at retrieving the sign at lower layers but there is a better correlation between the sign of the polarization and the radiation field tensors at higher layers. \par

The last finding highlights that, although simple and useful, the underlying assumption, namely that the scattering polarization is purely determined by the radiation field, is inherently incomplete. Stronger structured magnetic fields can break this correlation via the Hanle effect.
\citet{DelPinoAleman2018} also showed that gradients of the macroscopic velocity (especially its horizontal component) plays a very important role in the scattering polarization of the Sr~{\sc i} line, as they significantly contribute to the lack of axial symmetry. Consequently, the fundamental limitation of the reconstruction method is set by the assumption that linear polarization is exclusively proportional to the radiation field tensor $J^2_2$.
\par

The influence of the Zeeman effect on the systematic error due to its impact on the linear polarization is likely negligible. First, it was not considered in the simulation and, therefore, the simulated polarization amplitudes are not affected. Secondly, the broad pre-filter of the observation implies a spectral averaging over a quite symmetric line profile, which is expected for the quiet Sun photosphere with relatively small plasma velocities, due to the transverse Zeeman effect leads to a cancellation and to a very small linear polarization amplitude.

\par

Despite its limitations, using the reconstruction method from \citet{Zeuner2020} on Sr~{\sc i} 4607\,\AA \ line images provides a reasonable estimate of scattering polarization patterns and a basis for comparing linear polarization strengths. This is especially useful when data from different sources are compared statistically, for example even when the solar scenes under study are quite different.
\par

For future investigations with high spatial resolution and enhanced polarimetric sensitivity, as those that will be possible with the instrumentation of the 4\,m Daniel K. Inouye Solar Telescope \citep[DKIST, e.g.,][]{Rimmele2022} 
or of the European Solar Telescope \citep[EST, ][]{EST2022}, this study poses a few interesting questions to the physics of scattering polarization beyond magnetic field determination. 
One example is the mismatch of signs of the radiation field tensor and the polarization due to velocity gradients, collisions, or to a structured magnetic field, indicative of the Hanle effect \citep{Zeuner2022}. \par

With increasing numbers of resolved disk center scattering observations, e.g. with DKIST, it will be possible to narrow down the unresolved magnetic field component with more accuracy. However, the noise requirements for such direct observations are very challenging. \citet{DelPinoAleman2021} showed an example with the spectrograph instrument ViSP, where a noise level of 4$\cdot 10^{-4}$ for 2\,s integration is assumed. Then the signal-to-noise ratio is large enough to make the polarimetric structures visible. But since this is a too optimistic assumption (those numbers are based on the predicted performance of the observation planning tool), in a realistic scenario one will need much longer integration times to achieve this noise level. 
In this case the reconstruction method offers a solution to extract scattering signals dominated by noise, making them suitable for spatially resolved Hanle interpretations.

\section*{Acknowledgments}
 We express our gratitude to the referee for providing valuable suggestions that have enhanced the clarity of the paper. The Fast Solar Polarimeter project is funded by the Max Planck Society (MPG) and by the European Commission, grant No. 312495 (SOLARNET). This project has received funding from the European Research Council (ERC) under the European Union's Horizon 2020 research and innovation program (grant agreement No. 695075). F. Z. acknowledges the International Max Planck Research School for Solar System Science and the funding received from the Swiss National Science Foundation under grant number 200020\_213147. T. P. A.'s participation in the publication is part of the Project RYC2021-034006-I, funded by MICIN/AEI/10.13039/501100011033, and the European Union “NextGenerationEU”/RTRP.
J. T. B. acknowledges the funding received from the European Research Council (ERC) under the European Union's Horizon 2020 research and innovation programme (ERC Advanced Grant agreement No. 742265). This research has made use of NASA's Astrophysics Data System Bibliographic Services.
The authors declare that they have no competing financial interests.

\vspace{5mm}
\facility{Dunn (FSPII)}

\software{PORTA \citep{Stepan2013},
          astropy \citep{astropy},  
          scikit-learn \citep{scikit-learn}, 
          matplotlib \citep{Hunter:2007},
          numpy \citep{numpy}
          }

\bibliography{references}{}
\bibliographystyle{aasjournal}



\end{document}